\pdfoutput=1
\documentclass[11pt]{article}
\usepackage{amsmath}

\usepackage{mathrsfs}
\usepackage{braket}
\usepackage{ascmac}
\usepackage{bm}
\usepackage{graphics}
\usepackage{graphicx}
\usepackage[colorlinks=true,urlcolor=blue,anchorcolor=black,citecolor=blue,linkcolor=black,filecolor=black,menucolor=black,pagecolor=black,linktocpage=true,pdfproducer=medialab,pdfa=true]{hyperref}
\usepackage[at]{easylist}
\usepackage{here}
\usepackage{cite}
\usepackage{amssymb}
\usepackage{mhchem}
\usepackage{physics}
\usepackage{authblk}
\usepackage{indentfirst}
\usepackage[subrefformat=parens]{subcaption}
\usepackage[top=30truemm,bottom=30truemm,left=20truemm,right=20truemm]{geometry}
\usepackage{color}
\usepackage[absolute,overlay]{textpos}

\usepackage{pagecolor}
\definecolor{darkpastelgreen}{rgb}{0.01, 0.75, 0.24}

\makeatletter
\@addtoreset{equation}{section}

\renewcommand{\maketitle}{
  \begin{center}
    {\Large\bfseries\@title} \\  
    \vspace{2cm}
    {\@author} \\          
    \vspace{3.5cm}
    \@thanks 
  \end{center}
}

\makeatother

\title{Rotating wormholes in five dimensions with equal angular momenta: large asymmetry regime}
\date{}
\author[$a$]{Keiya Uemichi \thanks{\href{mailto:uemichi.keiya.j4@s.mail.nagoya-u.ac.jp}{uemichi.keiya.j4@s.mail.nagoya-u.ac.jp}}}
\author[$b$]{Yasutaka Koga}
\author[$c$]{Daiki Saito}
\author[$a,d$]{Chul-Moon Yoo}
\author[$e$]{Daisuke Yoshida}
\affil[$a$]{\small Division of Science, Graduate School of Science, Nagoya University, Nagoya 464-8602, Japan}
\affil[$b$]{\small Department of Information and Computer Science, Osaka Institute of Technology, \protect\\ Hirakata, Osaka 573-0196, Japan}
\affil[$c$]{\small Department of Physics, Kyoto University, Kyoto 606-8502, Japan}
\affil[$d$]{\small Kobayashi-Maskawa Institute for the Origin of Particles and the Universe (KMI), 
\protect\\ Nagoya University, Nagoya 464-8602, Japan}
\affil[$e$]{\small Department of Mathematics, Nagoya University, Nagoya 464-8602, Japan}

\begin{document}
\begin{flushright}
NU-QG-14,
KUNS-3095
\end{flushright}

\maketitle
\thispagestyle{empty}

\begin{abstract}
We clarify the relationship between rotation and the energy condition for stationary rotating wormhole solutions of the Einstein equations coupled to a phantom field in five-dimensional spacetime with equal angular momenta, particularly with large asymmetry between the two sides.
It was shown by Dzhunushaliev et al. that
the violation of the null energy condition can become arbitrarily small due to rotation.
We find that the degree of violation of the null energy condition is essentially determined by the angular momentum and shows little dependence on asymmetry, that is, the mass difference between the two asymptotic regions. We also discuss the relation between the wormhole spacetime and the Myers--Perry black hole. We find that the geometry asymptotes to the extremal Myers--Perry spacetime in the limit of large angular momentum, while the non-extremal black hole geometry cannot be reproduced in any limit.
\end{abstract}
\pagebreak
\tableofcontents

\section{Introduction}
A wormhole is a structure that connects two separate regions of spacetime, either two distinct universes or distant points within the same universe.
In many cases, the minimal-area surface on a spatial hypersurface connecting the two regions is referred to as the throat, and wormhole geometries are often characterized by the presence of a throat.
The earliest idea of a wormhole-like structure appeared in the work of Flamm in 1916~\cite{Flamm:1916, Flamm2015-yp}, shortly after the formulation of general relativity. 
Later, Einstein and Rosen proposed what is now known as the Einstein--Rosen bridge, interpreted as a ``bridge'' connecting two separate sheets of spacetime~\cite{einstein1935particle, Visser:1995cc}.
However, the throat closes in time, rendering it non-traversable for any particle attempting to pass through.

Motivated by the idea of constructing traversable wormholes, later studies introduced solutions supported by a phantom scalar field, such as the Ellis--Bronnikov wormhole~\cite{ellis1973ether, bronnikov1973scalar} and its asymmetric~\cite{Armendariz-Picon:2002gjc} and multi-sheet~\cite{Makita:2025bao} generalizations. 
The geometry of more general traversable wormholes is studied by Morris and Thorne~\cite{Morris:1988cz, Morris:1988tu}.
These wormhole spacetimes admit a traversable throat but inevitably require exotic matter that violates the null energy condition (NEC) at least near the throat~\cite{Morris:1988cz, Visser:1995cc}. 
While the violation of the energy condition may seem problematic, it is known that certain quantum effects, such as the Casimir effect, can lead to NEC violation in realistic physical settings~\cite{Lamoreaux:1996wh}.
For instance, Maldacena, Milekhin, and Popov~\cite{Maldacena:2018gjk} have considered traversable wormholes supported by quantum fields that make use of such effects. An application of this construction method using effective field theory corrections to gravity was further explored in Ref.~\cite{Kanai:2025zgq}.
Other aspects, such as the dynamical formation of wormholes~\cite{Hayward:2004wm,Koga:2025bqw,Koga:2026jvs} have also been discussed. These advances open up new possibilities for constructing realistic wormhole solutions 
in semiclassical or fully quantum gravitational theories. 

Building on these developments, another promising direction has been to relax the requirement for appreciable violation of the energy condition by considering more general classes of wormhole solutions.
In particular, rotating wormhole spacetimes, which possess axial rather than spherical symmetry, have been studied as candidates where the amount of exotic matter might be reduced. 
From an astrophysical standpoint, this is a reasonable extension, as astrophysical objects rotate in general. 
One of the earliest studies of rotating wormholes was conducted in Ref.~\cite{Teo:1998dp}, where a stationary, axisymmetric wormhole solution supported by exotic matter was investigated.
While this solution was not derived from a specific matter model, it provided a useful starting point for understanding the geometry and potential observational features~\cite{Jusufi:2017mav} of rotating wormholes.

Subsequently, Kashargin and Sushkov~\cite{Kashargin:2007mm, Kashargin:2008pk} investigated rotating wormhole solutions in four dimensions with a phantom scalar field by approximately solving the Einstein equations under the assumption of slow rotation, and demonstrated that increasing angular momentum reduces the amount of violation of the energy condition. 
However, their analytic treatment was limited to the slow-rotation regime, and extending the analysis to arbitrary angular momentum remains technically challenging. 
This is because introducing rotation breaks spherical symmetry.
As a result, the Einstein equations no longer reduce to ordinary differential equations (ODEs), but instead take the form of partial differential equations, which are considerably more difficult to solve. 
Although there have been some efforts in the numerical construction of rotating wormhole solutions~\cite{Chew:2016epf, Kleihaus:2014dla} 
and their stability analyses~\cite{Khoo:2024yeh, Azad:2024axu}, their general properties are still unknown, particularly in the regime of rapid rotation.

To evade this difficulty, 
Dzhunushaliev et al.~\cite{Dzhunushaliev:2013jja} considered wormholes supported by a phantom scalar field in five-dimensional spacetime with equal angular momenta. 
In five dimensions, unlike in four, there are two independent rotation planes, and by setting the two angular momenta equal, the resulting geometry preserves a higher symmetry than the case of non-equal angular momenta. 
This enhanced symmetry effectively reduces the problem to one involving only the radial coordinate, allowing the Einstein equations to be written as a system of ODEs despite the presence of rotation.
This simplification enables the numerical construction of rotating wormhole solutions beyond the slow-rotation regime. 
The results in Ref.~\cite{Dzhunushaliev:2013jja} suggest that, in the rapid-rotation regime, the wormhole solution asymptotically approaches the extremal Myers--Perry solution~\cite{Myers:1986un, Myers:2011yc}. 
Moreover, Dzhunushaliev et al.~\cite{Dzhunushaliev:2013jja} also initiated the stability analysis of rotating wormholes. 
We follow the strategy adopted in Ref.~\cite{Dzhunushaliev:2013jja} and investigate the five-dimensional wormhole solutions with equal angular momenta, including highly spinning cases.

In Ref.~\cite{Dzhunushaliev:2013jja}, the asymmetry parameter, which corresponds to the mass difference between the two asymptotically flat regions connected by the throat, was not examined in detail. 
As a result, it remains unclear whether the violation of the energy condition is generally reduced by the effect of rotation. 
In addition, since the solution space has not been fully explored in previous works, the location of the black hole limit in the phase diagrams of wormhole solutions has not been fully clarified.
In this work, we investigate rotating wormhole solutions of the Einstein equations coupled to a phantom scalar field with equal angular momenta in five dimensions by extending the parameter space, with particular emphasis on the asymmetry parameter and its possible implications for the relation between wormhole and black hole solutions.

This paper is organized as follows.
In Sections~\ref{sec:action} and \ref{sec:AsyBDC}, we provide the necessary setup to solve the Einstein equations under our ansatz. 
In particular, we discuss the degrees of freedom associated with asymptotic behaviors and boundary conditions.
In Section~\ref{sec:num_pro}, we present the numerical procedure used to solve the equations of motion. 
The results of the numerical calculations, the dependence of the NEC violation on the parameters, and the relation with the Myers--Perry black hole solution are discussed in Section~\ref{sec:structure}. 
Finally, Section~\ref{sec:conclusion} summarizes the paper.

\section{Action and field equations}
\label{sec:action}

In this section, we explain our setup, following Ref.~\cite{Dzhunushaliev:2013jja}. 
We consider five-dimensional
Einstein gravity coupled to a phantom massless scalar field $\phi$, where the action is given by 
\footnote{The constant $G$ is related to the Newtonian constant $G_N$ by $G_N=\frac{8}{3\pi}G$.}
 \begin{equation}
\label{eq_Einstein_action}
    S=\int d^5x \sqrt{-g} \left[ \frac{1}{16\pi G} R+\frac{1}{2}\partial_\mu \phi \partial^\mu \phi \right].
\end{equation} 
The equations of motion are given by 
\begin{eqnarray}
    \label{eq_Einstein}
    &&R_{\mu \nu}-\frac{1}{2}g_{\mu \nu}R=8\pi G T_{\mu \nu}, \\
    \label{eq_KG}
    &&\nabla^\mu\nabla_\mu\phi=0,  
\end{eqnarray}
where the stress-energy tensor $T_{\mu \nu}$ is given by 
\begin{align}
\label{eq_stress_enegy}
    T_{\mu \nu}=\frac{1}{2} g_{\mu \nu} \partial^{\rho}\phi\partial_{\rho}\phi-\partial_\mu\phi \partial_\nu \phi. 
\end{align}
From Eqs.~\eqref{eq_Einstein} and \eqref{eq_stress_enegy}, 
we obtain 
\begin{align}
\label{eq_Einstein_scalar}
    R_{\mu \nu}=-8\pi G \partial_\mu \phi \partial_\nu \phi.
\end{align}

\subsection{Metric ansatz}
Let us discuss the metric ansatz from the viewpoint of symmetry.
First, let us briefly mention the metric ansatz for a static, spherically symmetric wormhole, not a rotating one, as the first step. 
The metric ansatz for static, spherically symmetric wormholes in five dimensions can be given by
\begin{align}
\label{ansatz_sta_WH}
    ds^2&=-U_0(l)dt^2+U_1(l)(dl^2+h(l)d \Omega^2_3) 
\end{align}
with
\begin{equation}
    h(l)=l^2+r^2_0, \label{def h}
\end{equation}
where $d\Omega_3^2$ is the metric of the unit three-sphere, and $r_0$ is a constant which characterizes the scale of this geometry. 
The range of the radial coordinate $l$ is defined as $-\infty < l <+\infty$, and $r_0$ gives the throat radius when the spacetime is symmetric under the sign reversal of the coordinate $l$. 
Let us express $d \Omega^2_3$ by using three one-forms $\sigma_i$ $(i=1,2,3)$ described by the Hopf coordinates $\ \theta,\varphi$, and $\psi$ with $0\leq \theta \leq \pi/2$, $0\leq \varphi <2\pi$, and $0 \leq \psi < 2\pi$ as follows: 
\begin{align}
\label{metic_S3}
  d \Omega^2_3&=\frac{1}{4} \left( \sigma_1^2+\sigma_2^2+\sigma_3^2\right)
  =d\theta^2 + \sin^2\theta d\varphi^2 + \cos^2 \theta d \psi^2 ,\\
  \sigma_1&=2\cos{(\psi+\varphi)} d\theta + 2\sin{(\psi+\varphi)} \sin{\theta}\cos{\theta} (d\psi-d\varphi),\notag\\
  \sigma_2&=-2\sin{(\psi+\varphi)} d\theta + 2\cos{(\psi+\varphi)} \sin{\theta}\cos{\theta} (d\psi-d\varphi),\notag\\
  \sigma_3&=d\psi+d\varphi + (\cos^2{\theta}-\sin^2{\theta}) (d\psi-d\varphi).\notag
\end{align}
The metric~\eqref{ansatz_sta_WH} has a cohomogeneity-1 symmetry, that is, all the submanifolds given by fixing the value of $l$ are homogeneous
admitting $R_{t} \times SO(4) $. 

The metric ansatz for a rotating wormhole, which is our main focus in this paper, can then be obtained by 
introducing the angular velocity $\omega(l)$ and squashing the $S^3$ as follows:
\begin{align}
    ds^2=-U_0(l)dt^2+U_1(l)\left(dl^2+ \frac{h(l)}{4}\left(\sigma_1^2+\sigma_2^2 \right) \right) + U_2(l) \frac{h(l)}{4} \left(\sigma_3-2\omega(l) dt \right)^2, 
\end{align}
where the spacetime remains cohomogeneity-1 and admits the reduced symmetry $R_{t} \times SU(2) \times U(1)$. 
We can rewrite this metric
as follows: 
\begin{align}
\label{ansatz_rot_WH}
    ds^2&=-e^{2a}dt^2+pe^{-q}dl^2+phe^{-q}d\theta^2 +phe^{q-2a}[\sin^2\theta(d\varphi-\omega dt)^2+\cos^2\theta(d\psi-\omega dt)^2]\notag\\
    &+ph (e^{-q}-e^{q-2a})\sin^2\theta\cos^2\theta(d\psi-d\varphi)^2,
\end{align}
where we have introduced three functional degrees of freedom, $a$, $p$, and $q$, instead of $U_{0}$, $U_{1}$, and $U_{2}$, which are defined by $U_0(l)=e^{2a(l)}$, $U_1(l)=p(l)e^{-q(l)}$ and $U_2(l)=p(l)e^{q(l)-2a(l)}$. 
Due to the cohomogeneity-1 nature of the metric, the scalar field $\phi$ can also be assumed to be a function of $l$:
\begin{align}
    \phi&=\phi(l).
\end{align}

\subsection{Equations of motion}
\label{sec:eom}

In this subsection, we rearrange
the equations of motion 
under the metric and scalar field ansatz explained above.
First, the scalar field equation~\eqref{eq_KG} can be expressed as 
\begin{equation}
\label{eq_scal_pre}
    \left( p\sqrt{h^3} \phi'\right)'=0,
\end{equation}
where the prime ($'$) denotes the differentiation with respect to $l$.
By integrating this equation, we obtain the expression of $\phi'$ as 
\begin{equation}
\label{eq_scal}
    \phi'=\frac{Q}{p\sqrt{h^3}}, 
\end{equation}
where $Q$ is the ``scalar charge" introduced as an integration constant. 
We note that if $Q=0$, $T_{\mu\nu}$ given by Eq.~\eqref{eq_stress_enegy} vanishes and there is no contribution of the scalar field to the geometry. 

From the Einstein equations~\eqref{eq_Einstein_scalar}, we obtain
\begin{equation}
\label{eq_Einstein_simple}
    R_{\mu \nu}=0~~{\rm for} ~~ (\mu,\nu) \ne (l,l)
\end{equation}
 because $\phi$ depends only on $l$. 
From $R^t_\varphi=0$, we obtain 
\begin{equation}
\label{eq_Einstein_Rtphi}
    \left(h^{5/2}p^2 e^{q-4a}\omega'\right)'=0.
\end{equation}
By integrating this equation, we find
\begin{equation}
\label{eq_Einstein_omega}
    \omega'=\frac{c_\omega}{h^{5/2}p^2e^{q-4a}}
\end{equation}
with an integral constant $c_{\omega}$. 
Moreover, incorporating with Eq.~\eqref{eq_Einstein_omega}, 
$R^t_t=0$ is reduced to the second-order differential equation of $a$:\\
\begin{equation}
\label{eq_Einstein_a}
    a''+\left(\frac{p'}{p}+\frac{3h'}{2h}\right)a'-\frac{c_\omega^2 e^{4a-q}}{2p^3 h^4}=0.
\end{equation}
Similarly, incorporating with Eqs.~\eqref{eq_Einstein_omega} and \eqref{eq_Einstein_a}, $R^\varphi_\psi=0$ is also reduced to the second-order differential equation of $q$:
 \begin{equation}
 \label{eq_Einstein_q}
   q''+\left(\frac{p'}{p}+\frac{3h'}{2h}\right)q'-\frac{4(e^{2(q-a)}-1)}{h}=0.
 \end{equation}
 \\
 Combining this equation with $R^\theta_\theta=0$, we obtain the differential equation for $p$:
\begin{align}
\label{eq_Einstein_p}
&p''+\frac{5h'}{2h}p'+\frac{h'^2-2h(4-h'')}{2h^2}p=0.
\end{align}
Using the concrete expression of $h(l)$ given by Eq.~\eqref{def h}, the general solution for this second-order differential equation is given by
\begin{align}
\label{eq_sol_p}
     p(l)=c_1\frac{l}{\sqrt{l^2+r_0^2}}+c_2\frac{l^2+r_0^2/2}{l^2+r_0^2}
\end{align}
with two integration constants $c_1$ and $c_2$. Since $p>0$ must be satisfied in any region, we require $
\lim_{l \to \pm \infty} p = c_2 \pm c_1 >0$. 

From $R_{ll}=-8\pi G\partial_l\phi\partial_l\phi$, we obtain
\begin{align} 
    -8\pi G \frac{Q^2}{h^3p^2}=& -\frac{3 p^{\prime \prime}}{2 p}+\frac{q^{\prime \prime}}{2}-\frac{3 h^{\prime \prime}}{2 h}+\frac{3 h^{\prime 2}}{4 h^2}+\frac{3 p^{\prime 2}}{2 p^2}-2 a^{\prime 2}-\frac{q^{\prime 2}}{2}+a^{\prime} q^{\prime}\notag \\
     & -\frac{p^{\prime}}{p}\left(\frac{q^{\prime}}{2}-a^{\prime}\right)-\frac{h^{\prime}}{4 h}\left(q^{\prime}+\frac{3 p^{\prime}}{p}-4 a^{\prime}\right)+\frac{1}{2} e^{q-4 a} h p \omega^{\prime 2}.
\end{align}
Here, we have used Eq.~\eqref{eq_scal}.
Eliminating $\omega'$, $q''$, and  $p''$ by using Eqs.~\eqref{eq_Einstein_omega}, \eqref{eq_Einstein_q}, and \eqref{eq_Einstein_p}, 
respectively, we find
\begin{align}
\label{eq_constr}
    4 \pi G Q^2=& -\frac{e^{4 a-q} c_{\omega}^2}{4 hp}+\frac{h^3}{4}\left[\left(4 a'^2-2 a' q'+q'^2\right) p^2-2\left(a'-q'\right) p p'-3 p'^2 \right]\notag\\
    &+h^2 p^2 \left(4-e^{2(q-a)}\right)+\frac{1}{2}h'h^2 p\left[p\left(q'-a'\right) -3p' \right]-\frac{3}{4}h'^2 h p^2.
\end{align}
Since this equation does not include the second derivative for $l$, it can be regarded as a constraint equation. 
In practice, we fix the value of $Q$ through Eq.~\eqref{eq_constr} by choosing the boundary conditions for $a$ and $q$. 
The functional forms of $a$ and $q$ are given by solving Eqs.~\eqref{eq_Einstein_a} and \eqref{eq_Einstein_q}. 
Once $a$ and $q$ are determined, $\omega$ can be derived by solving the first-order differential equation Eq.~\eqref{eq_Einstein_omega}.\\

Notably, the parameters $(c_1, c_2)$ represent degrees of freedom in the coordinate transformation. 
For instance, to simplify the expression, let us consider a transformation resulting in $(c_1, c_2) = (0, 1)$. 
Such a transformation is realized by
\begin{align}
    \Tilde{l}=\pm  \left(\frac{\sqrt{c_2+c_1}+\sqrt{c_2-c_1}}{2}l+\frac{\sqrt{c_2+c_1}-\sqrt{c_2-c_1}}{2} \sqrt{l^2+r_0^2}\right).
\end{align}
After the transformation from $l$ to $\Tilde{l}$, 
$ph$ and $p dl^2$
take the form:   
\begin{align}
\label{eq_pl}
    p(l)dl^2&=\tilde p(\tilde l)d{\Tilde l}^2:=\frac{{\Tilde l}^2+{\Tilde r}_0^2/2}{{\Tilde l}^2+{\Tilde r}_0^2}d{\Tilde l}^2, \\
    p(l)h(l) &= c_1 l \sqrt{l^2 + r_0^2} + c_2 \left(l^2 + r_0^2/2\right) = {\Tilde{l}}^2 + {\Tilde{r}}_0^2/2=\Tilde p(\Tilde l)\Tilde h(\Tilde l),
\end{align}
where 
\begin{align}
\Tilde h (\Tilde l)&: ={\Tilde l}^2+{\Tilde r}_0^2, \\
 {\Tilde{r}}_0^2 &: = \sqrt{c^2_2 - c^2_1} r_{0}^2.
\end{align}
In Eq.~\eqref{eq_pl}, the right-hand side corresponds to the functional form of $p$ for $(c_1, c_2) = (0, 1)$, expressed in the new coordinate $\Tilde{l}$ and ${\Tilde{r}}_0$. 
Thus, arbitrary values of $(c_1, c_2)$ can be transformed to $(c_1, c_2) = (0, 1)$.
Using this property, we will fix $(c_1, c_2) = (0, 1)$ in what follows.

\section{Boundary conditions and asymptotic behaviors}
\label{sec:AsyBDC}

Let us discuss boundary conditions 
to solve equations~\eqref{eq_scal}, \eqref{eq_Einstein_omega},  \eqref{eq_Einstein_a}, and \eqref{eq_Einstein_q}. 
In the following part of this paper, we fix $r_0=1/2$ for the numerical calculations.
Since we are focusing on the massless scalar system, the constant shift of the value of the scalar field $\phi$ can be arbitrary, and we set 
\begin{align}
    &\phi(+\infty)=0. 
\end{align}
As for the boundary condition for $\omega$ to solve 
Eq.~\eqref{eq_Einstein_omega}, 
without loss of generality,
we fix
\begin{align}
    &\omega(+\infty)=0,
\end{align}
because a constant shift of $\omega$ ($\omega \to \omega + C$) can be absorbed by a time-dependent transformation
of the angular coordinates ($\varphi \to \varphi - Ct$ and $\psi \to \psi - Ct$).

To solve the two second-order differential equations \eqref{eq_Einstein_a} and \eqref{eq_Einstein_q}, we need four boundary conditions. 
Moreover, since $c_\omega$ is an arbitrary constant, we need to fix five parameter values in total. 
As for the boundary condition, we require that the functions $a$ and $q$ 
can be expanded in negative powers of $l$ at $l\to \pm \infty$. 
As we will see in the following, this condition 
reduces the number of required parameters, and 
ensures the asymptotic flatness at $l\to \pm \infty$.  

Substituting the expanded form of the functions $a$ and $q$ into equations~\eqref{eq_Einstein_a} and~\eqref{eq_Einstein_q}, we obtain the following asymptotic forms:
\begin{align}
\label{asymp_aq}
 a(l)=a_{\pm \infty}+\sum_{i=1}^8 \frac{a^{\pm}_i}{l^i}+O(l^{-9}),\quad q(l)=q_{\pm \infty}+\sum_{i=1}^8 \frac{q^{\pm}_i}{l^i}+O(l^{-9})\notag
\end{align}
with
\begin{align}
& a_4^{\pm}=-\frac{a_2^{\pm} r_0^2}{2}, \notag\\
& a_6^{\pm}=\frac{1}{48}\left(c_\omega^2 e^{3 a_{\pm \infty}}+14 a_2^{\pm} r_0^4\right), \notag\\
& a_8^{\pm}=-\frac{3 r_0^6}{16} a_2^{\pm}+\frac{c_\omega^2 e^{3 a_{\pm \infty}}}{32}\left(a_2^{\pm}-r_0^2\right), \notag\\
& a_{2 n-1}^{\pm}=0 \quad (n \in \mathbb{N}), \notag\\
& q_{\pm\infty}=a_{\pm\infty}, \notag\\
& q_2^{\pm}=a_2^{\pm}, \notag\\
& q_6^{\pm}=-\frac{5 r_0^4}{24} a_2^{\pm}-r_0^2 q_4^{\pm}-\frac{c_\omega^2 e^{3 a_{\pm \infty}}}{96}, \notag\\
& q_8^{\pm}=\frac{1}{80}\left[19 r_0^6 a_2^{\pm}+68 r_0^4 q_4^{\pm}+4 r_0^4\left(a_2^{\pm}\right)^2+16\left(q_4^{\pm}\right)^2+16 r_0^2 a_2^{\pm} q_4^{\pm}\right]+\frac{c_{\omega}^2 e^{3 a_{\pm \infty}}}{320}(5r_0^2-2a_2^{\pm}), \notag\\
& q_{2 n+1}^{\pm}=0 \quad (n \in \mathbb{N}).
\end{align}
The superscript $\pm$ indicates that the coefficients correspond to the expansions around $l \to \pm \infty$, respectively.
Focusing first on the case of $l \to +\infty$, the asymptotic expansion can be expressed in terms of three parameters, $a_{+\infty}$, $a_2^{+}$, and $q_4^{+}$.
Thus, our boundary conditions on one side eliminate one parameter degree of freedom in the general solution.
Since we impose the boundary conditions on both sides of $l\to \pm\infty$, we have the reduction of two parameter degrees of freedom in total. 
Consequently, the number of free parameters is reduced from five to three. 
For instance, we may choose the three independent parameters as $a_{+\infty}$, $a_{-\infty}$, and $c_{\omega}$. 

Additionally, substituting these expressions to Eq.~\eqref{eq_Einstein_omega}, we can obtain the asymptotic behaviors of $\omega$:
\begin{align}
\label{asymp_omega}
    \omega=\omega_{\pm \infty}\mp\frac{c_{\omega}}{4l^4}e^{3a_{\pm \infty}}+O(l^{-5}).
\end{align}

For each asymptotic region, we shall introduce the following coordinates: 
\begin{align}
\label{rela_transcoord}
    t_{\pm}&=e^{a_{\pm \infty} }t, \notag\\
    l_{\pm}&=e^{-\frac{1}{2}a_{\pm \infty}} l,\notag\\
    \varphi_{\pm}&=\varphi-\omega_{\pm \infty} t,\notag\\
    \psi_{\pm}&=\psi-\omega_{\pm \infty} t. 
\end{align}
Then, 
in the limit $l\rightarrow\pm\infty$, the metric \eqref{ansatz_rot_WH} can be written in the flat form
\begin{align}
    ds^2 \rightarrow -dt_{\pm}^2+dl_{\pm}^2+ l_{\pm}^2 ( d\theta^2 + \sin^2\theta d\varphi_{\pm}^2 + \cos^2\theta d\psi_{\pm}^2 ), 
\end{align}
which explicitly shows that our boundary conditions guarantee the asymptotic flatness. 
The angular part of this metric corresponds to the three-sphere metric \eqref{metic_S3}.
Since the value of $a_{+\infty}$ or $a_{-\infty}$ can be absorbed into the redefinition of the time coordinate as
$t \to \Tilde{t}=e^{a_{+\infty}}t$ or $t \to \Tilde{t}=e^{a_{-\infty}}t$,
we can set $a_{+\infty} = 0$ without a loss of generality.
Then, regarding $a_{-\infty}$ as a free parameter, we have two parameter degrees of freedom together with $c_{\omega}$. 
That is, we obtain a 2-parameter $(c_{\omega}, a_{-\infty})$ family of solutions.

Let us express the parameters $a_2^{\pm}$ and $q_4^{\pm}$ in terms of the global charges.
Since the wormhole possesses two asymptotically flat regions, we define 
the mass $M_\pm$ and the angular momentum $J_\pm$ in the asymptotic region $ l \to \pm \infty $. 
These quantities are defined 
through the $t_{\pm} t_{\pm}$ and $t_{\pm} \varphi_{\pm}$ components of the metric as follows:
\begin{align}
    g_{t_{\pm}t_{\pm}} &=  - \left( 1-\frac{8G M_\pm}{3 \pi l^2_{\pm}} \right)+O\left(l_{\pm}^{-3}\right), \label{asympcondM} \\
    g_{t_{\pm} \varphi_{\pm}} &= -  \frac{4G J_\pm \sin^2{\theta}}{\pi l^2_{\pm}} +O\left(l_{\pm}^{-3}\right).
    \label{asympcondJ}
\end{align} 

Comparing Eq.~\eqref{asympcondM} to Eq.~\eqref{asymp_aq}, we find the relation between $a_2^{\pm}$ and $M_{\pm}$ as
\begin{align}
\label{rela_M_a2}
    M_{\pm}=-\frac{3\pi e^{-a_{\pm \infty}}}{4G}a_2^{\pm}.
\end{align}
Comparing Eq.~\eqref{asympcondJ} to Eq.~\eqref{asymp_omega}, we find the relation between $c_{\omega}$ and $J_{\pm}$ as
\begin{align}
\label{rela_J_comega}
    -J_+=J_-=\frac{\pi}{16G} c_{\omega}=:J.
\end{align}
Therefore, the angular momenta $J_+$ and $ J_-$ are equal to each other and proportional to $c_{\omega}$. 
The scalar charge $Q$ can be related to the parameter $q_4^{\pm}$, which appears in the asymptotic expansion, through the constraint equation~\eqref{eq_constr},
\begin{align}
\label{rela_Q_q4}
    Q^2=\frac{3}{2 \pi G}\qty[-q_4^{\pm}+\frac{r_0^4}{8}-\frac{r_0^2}{2} a_2^{\pm}+\frac{1}{2}\left(a_2^{\pm}\right)^2].
\end{align}

\section{Numerical procedure}
\label{sec:num_pro}
To obtain solutions for Eqs.~\eqref{eq_Einstein_a}, \eqref{eq_Einstein_q}, and \eqref{eq_constr}, we need to impose the boundary conditions discussed in Sec.~\ref{sec:AsyBDC} in both asymptotic ends. 
In this work, we employ a shooting method to solve this problem. 
It turns out that the system is unstable when we solve the differential equations towards an asymptotic end. 
Therefore, we solve the differential equations from both ends and impose smooth matching conditions at an intermediate point. 
In general, for solving the two independent second-order differential equations, we need to impose 4 matching conditions on the values of the two dependent variables and their derivatives. 
However, since there is a constraint equation \eqref{eq_constr}, one of the matching conditions will be automatically satisfied once the other three are imposed. 
Therefore, we have three independent matching conditions.  

To avoid treating the infinite value of the radial coordinate $l$, we introduce a coordinate $x$ defined by 
\begin{align}
    x=\frac{2}{\pi}\arctan{(l/r_0)},\quad x \in (-1,1).
\end{align}
By applying this transformation, the differential equations \eqref{eq_Einstein_a} and \eqref{eq_Einstein_q} can be rewritten as
\begin{align}
\label{eq_aq_x}
& \frac{d^2 a}{d x^2}+\frac{\pi}{2} \tan \left(\frac{\pi}{2} x\right) \frac{2 \tan ^2\left(\frac{\pi}{2} x\right)+3}{2 \tan ^2\left(\frac{\pi}{2} x\right)+1} \frac{d a}{d x}-\frac{\pi^2 c_\omega^2}{r_0^6} e^{4 a-q} \frac{\tan ^2\left(\frac{\pi}{2} x\right)+1}{\left(2 \tan ^2\left(\frac{\pi}{2} x\right)+1\right)^3}=0, \notag\\
& \frac{d^2 q}{d x^2}+\frac{\pi}{2} \tan \left(\frac{\pi}{2} x\right) \frac{2 \tan ^2\left(\frac{\pi}{2} x\right)+3}{2 \tan ^2\left(\frac{\pi}{2} x\right)+1} \frac{d q}{d x}-\pi^2\left(e^{2(q-a)}-1\right) \left(\tan ^2\left(\frac{\pi}{2} x\right)+1\right)=0.
\end{align}
Since the boundary conditions are imposed at $ x \to \pm 1 $, we examine their asymptotic behaviors 
for $x=\pm 1\mp \epsilon$ with $\epsilon\to 0$. 
From Eq.~\eqref{asymp_aq}, we find the following forms of the expansion in terms of the coordinate $x$: 
\begin{align}
    a(\pm 1\mp\epsilon)&
    =a_{\pm\infty}+\frac{\pi^2 a^{\pm}_2}{4r_0^2}\epsilon^2
    +\qty(\frac{\pi^4 a^{\pm}_4}{16r_0^4}+\frac{\pi^4 a^{\pm}_2}{24 r_0^2})\epsilon^4
    +\qty(\frac{\pi^6 a_6^{ \pm}}{64 r_0^6}+\frac{\pi^6 a_4^{ \pm}}{48 r_0^4}+\frac{17 \pi^6 a_2^{ \pm}}{2880 r_0^2}) \epsilon^6\notag\\
    &+\qty(\frac{\pi^8 a_8^{ \pm}}{256 r_0^8}+\frac{\pi^8 a_6^{\pm}}{128 r_0^6}+\frac{3 \pi^8 a_4^{ \pm}}{640 r_0^4}+\frac{31 \pi^8 a_2^{ \pm}}{40320 r_0^2})\epsilon^8
    +O(\epsilon^{10}),\notag\\
    q(\pm 1\mp\epsilon)&
    =q_{\pm\infty}+\frac{\pi^2 q^{\pm}_2}{4r_0^2}\epsilon^2
    +\qty(\frac{\pi^4 q^{\pm}_4}{16r_0^4}+\frac{\pi^4 q^{\pm}_2}{24 r_0^2})\epsilon^4
    +\qty(\frac{\pi^6 q_6^{ \pm}}{64 r_0^6}+\frac{\pi^6 q_4^{ \pm}}{48 r_0^4}+\frac{17 \pi^6 q_2^{ \pm}}{2880 r_0^2}) \epsilon^6\notag\\
    &+\qty(\frac{\pi^8 q_8^{ \pm}}{256 r_0^8}+\frac{\pi^8 q_6^{\pm}}{128 r_0^6}+\frac{3 \pi^8 q_4^{ \pm}}{640 r_0^4}+\frac{31 \pi^8 q_2^{ \pm}}{40320 r_0^2})\epsilon^8
    +O(\epsilon^{10}). 
\end{align}
As stated above, the system
has two independent degrees of freedom, represented by $ a_{-\infty} $ and $c_{\omega}$.
The other independent parameters $a_2^\pm$, $q_4^{\pm}$ in Eq.~\eqref{asymp_aq} can be replaced by $M_{\pm}$ and $Q$ by using Eqs.~\eqref{rela_M_a2} and \eqref{rela_Q_q4}.
Accordingly, $M_{\pm}$ and $Q$ are employed as shooting parameters in our numerical procedure in the following.

Here, let us summarize our procedure to obtain the two-parameter family of solutions. 
First, we start our procedure by considering the analytic static solution characterized by $a_{-\infty}$ 
with $c_\omega = 0$ (see Appendix~\ref{app:static_solution}). 
Gradually changing the value of $c_\omega$, we perform the iteration procedure of the shooting method for each value of $c_\omega$ to obtain the values of $(M_+, Q, M_-)$
imposing the matching condition at $x=0$. 
The iteration procedure of the shooting method involves integrating from both ends and adjusting $(M_+, Q, M_-)$ to minimize the differences in $\left(a, a', q'\right)$ at the intermediate point $x = 0$. 
The continuity of $q$ is guaranteed by the constraint Eq.~\eqref{eq_constr}. 
The initial trial values of $(M_+, Q, M_-)$ are taken from the previous convergent values for a slightly smaller value of $c_\omega$. 
By successively increasing $c_\omega$ in this way and determining $(M_+, Q, M_-)$ at each step, the corresponding solutions are constructed.
The above procedure is repeated for each value of $a_{-\infty}$. 

\section{Numerical solutions}
\label{sec:structure}
In this section, we present our numerical solutions.

\subsection{Area of wormholes}
First, we examine the area $A$ of the 3-dimensional cross section at fixed $t$ and $l$: 
\begin{align}
    \label{def_A}
    A&=\int_{0}^{\pi/2} d\theta \int_{0}^{2\pi} d\varphi \int_{0}^{2\pi} d\psi \sqrt{g^{(3)}}
    =2\pi^2\left[(ph)^3 e^{-(2a+q)} \right]^{1/2}, 
\end{align}
where $g^{(3)}$ is the determinant of the induced metric on the surface of constant $t$ and $l$. 
We define the location of the wormhole throat $l_{\text{th}}$ as the position where the cross-sectional area takes the minimum value, thus,
\begin{align}
 A'|_{l = l_{\text{th}}} = 0. 
\end{align}
We define, then, the throat area $A_{\text{th}}$ as 
\begin{align}
\label{def_Ath}
A_{\text{th}} = A|_{l=l_{\text{th}}}.
\end{align}

\begin{figure}[H]
    \centering
        \centering
        \includegraphics[width=6.3cm]{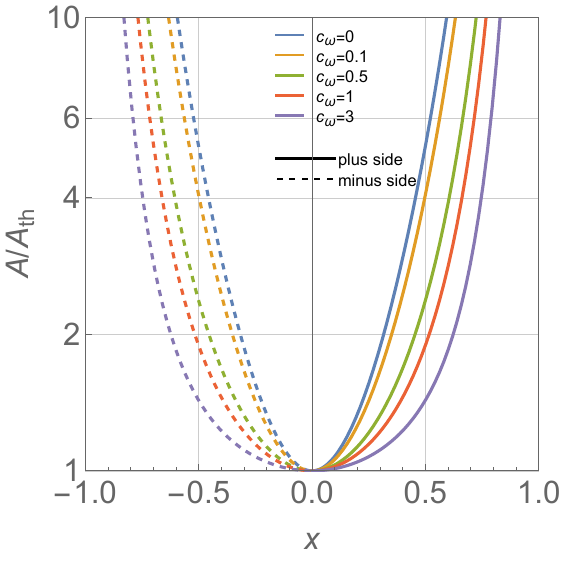}
        \includegraphics[width=6.3cm]{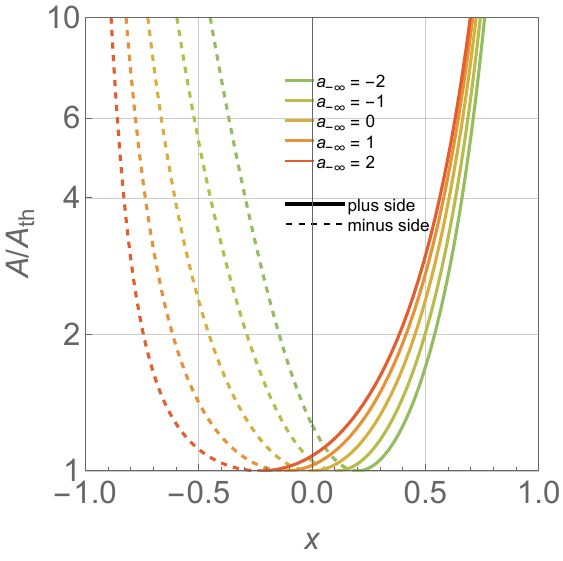}
         \caption{
               The cross-sectional area $A$ as a function of $x$. The left panel shows the results for fixed $a_{-\infty}=0$ and $c_{\omega}=0, 0.1, 0.5, 1, 3$, while the right panel shows those for fixed $c_{\omega}=0.5$ and $a_{-\infty}=-2, -1, 0, 1, 2$. 
               The solid and dashed curves describe the profiles in the region $l>l_{\rm th}$ and $l<l_{\rm th}$, respectively. 
               }
        \label{fig_xvsA}
\end{figure}

We plot the value of $A$ as a function of $x$ for each value of $c_\omega$ and $a_{-\infty}$ in Fig.~\ref{fig_xvsA}. 
The left and right panels in Fig.~\ref{fig_xvsA} show the behaviors of the area for the symmetric ($a_{+\infty}=a_{-\infty}=0$) and asymmetric ($a_{+\infty}=0$ and $a_{-\infty} \ne 0$) cases, respectively. 
As is expected, the figure in the left panel has the reflection symmetry with respect to $x=0$, while the figure in the right panel shows asymmetric features.
In the left panel, the structure of the throat is stretched along the $x$-direction
as $c_{\omega}$ increases.
In the right panel, the throat shifts
and the figure lacks reflection symmetry with respect to the throat for $a_{-\infty}\neq0$.

The value of $x$ is just a coordinate value, and does not necessarily reflect geometrical characteristics of the wormhole structure. 
Therefore, hereafter, we use the cross-sectional area $A$ to specify the radial location instead of $x$. 
In the same manner, $r_0$, which we set to $1/2$ in the numerical integration, is not the value of the proper length scale. 
Thus, we use the throat area $A_{\rm th}$, given by Eq.~\eqref{def_Ath}, as a normalization factor for dimensionful quantities.
The corresponding metric functions $a$, $q$, and $\omega$, expressed as functions of the area $A$, are shown in Appendix~\ref{App:metric_function}.

\subsection{Phase diagrams}
\label{sec:Phasediagram}
\begin{figure}[htbp]
        \centering
        \includegraphics[width=16cm]{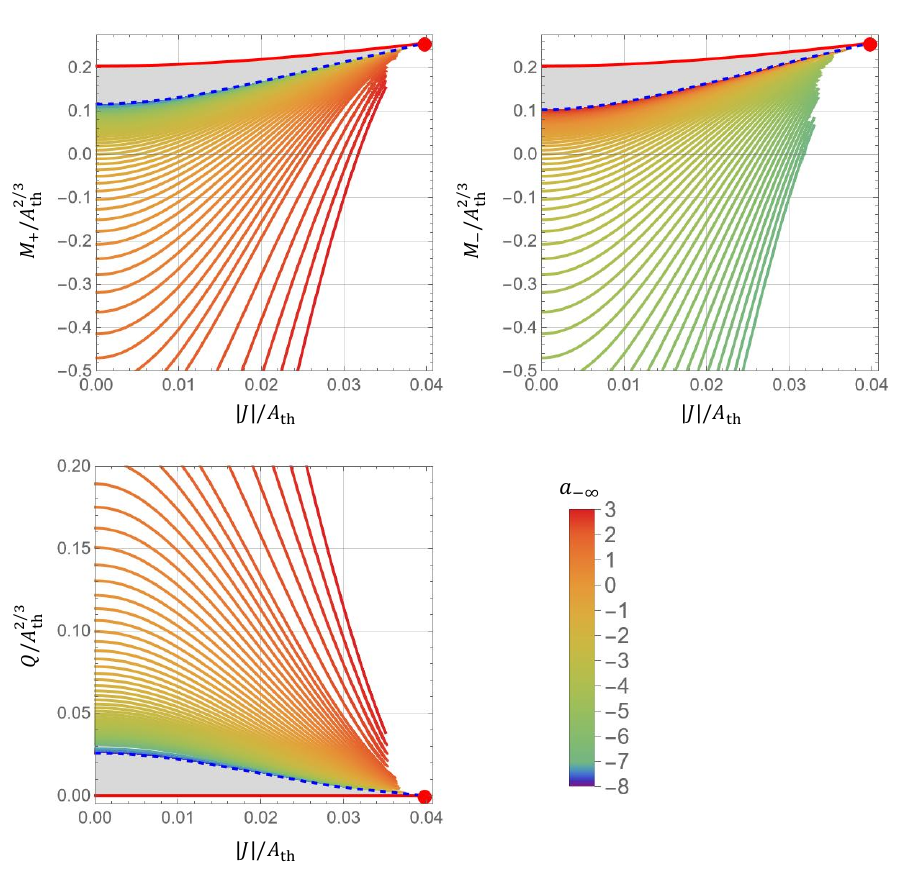}
        \caption{
        Phase diagrams showing the dependence of the mass for $l \to +\infty$, ~$M_{+}$ (upper left), the mass for $l \to -\infty$, ~$M_{-}$ (upper right), and the scalar charge~$Q$ (lower left) on the angular momentum~$J$.
        The values of $M_{+}, M_{-}, Q$ and $J$ are normalized by the area of the throat $A_{\text{th}}$ 
        (with $G=1$),
        with color variations corresponding to the value of $a_{-\infty}$ for fixed $r_0$. 
        The red curves correspond to the Myers--Perry solution, and these endpoints, the red points at $\abs{J}/A_{\text{th}} \sim 0.04$, indicate the extremal solutions.
        The blue-dashed curves represent the extrapolation toward the red points for $a_{-\infty}=-8$ (upper left and lower left) and $a_{-\infty}=3$ (upper right), converging to the limits $a_{-\infty}\to -\infty$ and $a_{-\infty}\to \infty$, respectively.
        The gray-shaded region shows the gap between the red curve and the blue-dashed curve.
        }
        \label{fig_JvsQorM}
\end{figure}

In this subsection, we show the phase diagrams of rotating wormhole solutions. We use global charges $J=\pi c_\omega/(16 G)$, $M_{+}$, $M_{-}$, and $Q$
as independent parameters characterizing the solutions, instead of $c_\omega$ and $a_{-\infty}$. 
First, we draw curves on the $|J|$-$M_+$, $\abs{J}$-$M_-$, and $|J|$-$Q$ planes each of which describes 
a one-parameter family of solutions characterized by the parameter $|J|$ for a fixed value of $a_{-\infty}$ in Fig.~\ref{fig_JvsQorM}.

Fig.~\ref{fig_JvsQorM} shows that both masses, $M_+$ and $M_-$, increase and the scalar charge, $Q$, decreases monotonically with increasing angular momentum $|J|$. 
Increasing $|J|$, 
around $|J|/A_{\text{th}}\simeq 0.035$, we observe poor convergence and eventually
failed to obtain a numerical solution.
As is pointed out in Ref.~\cite{Dzhunushaliev:2013jja}, it is notable that the values $M_{\pm}/A_{\text{th}}^{2/3}\simeq 0.256$, $|J|/A_{\text{th}}\simeq 0.0398$ and $Q=0$
indicated by the red points in the phase diagrams (Fig.~\ref{fig_JvsQorM}) correspond to those for the extremal Myers--Perry black hole with $A_{\rm th}$ replaced by the horizon area (see Eqs.~\eqref{exMP_mass} and \eqref{exMP_AM} in Appendix \ref{App:Rela_MPBH_WH_analy} for some analytic formulae). 
It seems that all the curves in the phase diagram approach the red point, and the point appears to be on the extrapolated extension of all the curves. 
That is, our numerical analyses suggest that
all sequences converge to the same point,
characterized by the mass values $M_{\pm}/A_{\text{th}}^{2/3}\simeq 0.256$ and the angular momentum $|J|/A_{\text{th}}\simeq 0.0398$, regardless of the value of $a_{-\infty}$.
This also may indicate the existence of a universal upper bound on $J$ for a fixed throat area.
It should also be noted that,
as this critical angular momentum is approached, the scalar charge tends to $Q=0$, indicating that the limit solution becomes a vacuum solution. 
In summary, all the results suggest that,
in the rapid rotation limit, the wormhole solution approaches the extremal Myers--Perry solution.
When $a_{-\infty}=0$, the two masses $M_{+}$ and $M_{-}$ exhibit identical behavior.
In contrast, for $a_{-\infty}\neq 0$, their behaviors differ, and the roles of $M_{+}$ and $M_{-}$ are interchanged depending on the sign of $a_{-\infty}$.

So far, we have been discussing the large $|J|$ or equivalently large $c_\omega$ limit. Here, let us check the $a_{-\infty}$ dependence of the solution, particularly focusing on the limit $a_{-\infty}\rightarrow \pm\infty$. 
That is, we investigate the solution beyond the specific value $a_{-\infty}=-1,0,1$
considered in \cite{Dzhunushaliev:2013jja}, 
and verify whether there exists a parameter region in which the geometry of the Myers--Perry black hole is realized.

As $a_{-\infty}$ decreases, the curves in 
the upper left panel in Fig.~\ref{fig_JvsQorM} 
shift upward, while those in 
the lower left panel in
Fig.~\ref{fig_JvsQorM}  
shift downward and finally seem to converge to the blue-dashed curve corresponding to the limit $a_{-\infty} \to -\infty$. 
In practice, the limit curve
is drawn by extrapolating the numerical result for $a_{-\infty} = -8$ so that it connects smoothly to the red point corresponding to the extremal Myers--Perry black hole. 
Given the convergent behavior of the numerical sequences, together with the exact solution for $J=0$ derived analytically in the limit $a_{-\infty}\to -\infty$ (Appendix~\ref{app:static_solution}), the curve for $a_{-\infty}=-8$ already provides a sufficiently accurate approximation for the limit curve.
The curves in the upper right panel in Fig.~\ref{fig_JvsQorM} 
shift upward as $a_{-\infty}$ increases as opposed to the upper left panel. 
They converge to the blue-dashed curves corresponding to the limit $a_{-\infty}\to+\infty$.
The red curve in Fig.~\ref{fig_JvsQorM} indicates the sequence of the non-extremal Myers--Perry black holes with equal angular momenta. 
The existence of the limit curve and the gray-shaded region in Fig.~\ref{fig_JvsQorM} 
imply that the values of $M_{\pm}/A_{\text{th}}^{2/3}$, $|J|/A_{\text{th}}$ and $Q=0$ 
corresponding to non-extremal Myers--Perry black holes cannot be realized by the wormhole solutions.  
Namely, 
even if we take the limit $a_{-\infty} \to -\infty$, the solutions do not approach the Myers--Perry black hole, leaving a gap corresponding to the gray-shaded region in the figure.

\subsection{Relation with the Myers--Perry black hole with equal angular momenta}
\label{sec:Rela_MPBH_Phasediag}

The results in the previous sections suggest that
the mass and angular momentum
of the wormhole solutions in the rapid rotation limit approach those of
the extremal Myers--Perry solution corresponding to the red point in Fig.~\ref{fig_JvsQorM}.
This would imply that the geometry of the wormhole solutions approaches the extremal Myers--Perry solution in the rapid rotation limit. 
If this implication is correct, it should be possible to express the extremal Myers--Perry solution in terms of the metric ansatz \eqref{ansatz_rot_WH}. 
We can explicitly show that the metric of the extremal Myers--Perry black hole is consistent with the metric ansatz \eqref{ansatz_rot_WH} and satisfies the equations described in this paper (see Appendix~\ref{App:Rela_MPBH_WH_analy}).

\begin{figure}[htbp]
  \centering
  \includegraphics[width=8cm]{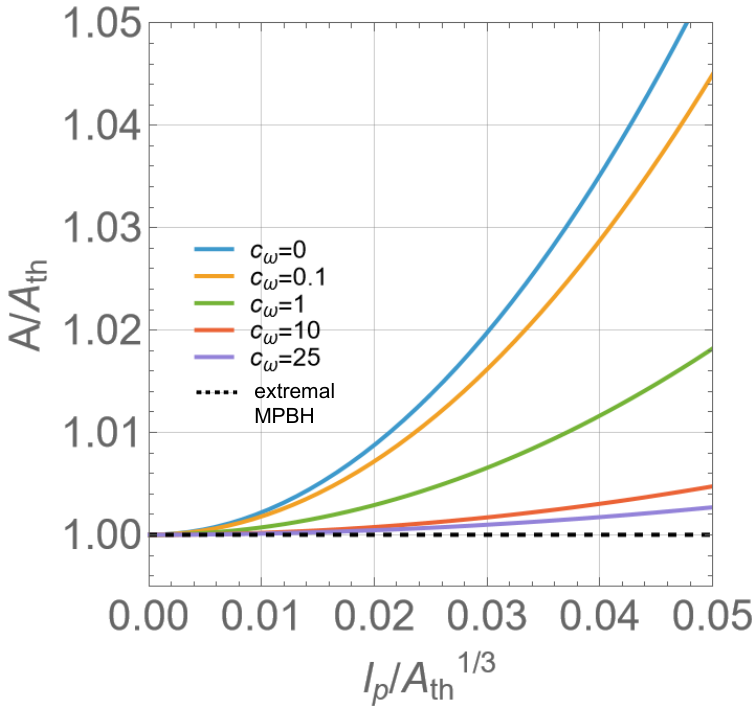}
  \caption{
  The dependence of the cross-sectional area $A$ on the proper length $l_{p}$ for symmetric case~($a_{-\infty}~=~0$).
  The solid curves are the result of the numerical integration for the cases of the wormhole, and the dotted horizontal line 
 corresponds to the case of the extremal Myers--Perry black hole.
  }
  \label{fig_Plot_lpvsA}
\end{figure}

Let us further check the geometrical properties in the rapid rotation limit. 
As is well known, in the extremal limit, the near-horizon region of the Myers--Perry black hole can be approximated by an infinitely long cylinder in a time slice. 
Therefore, we expect that the wormhole structure near the throat is stretched in the rapid rotation limit.  
We plot the cross-sectional area $A$ as a function of the radial proper length $l_p$ measured from the throat, defined by
\begin{align}
    l_p(l)=\int^{l}_{l_{\text{th}}} \sqrt{g_{ll}(l')} \, dl',
\end{align}
in Fig.~\ref{fig_Plot_lpvsA}.
We can find that, as the 
angular momentum increases, the change in the area as a function of $l_p$ becomes less noticeable,
indicating that the structure is stretched, as in the case of the extremal Myers--Perry black hole. 

\subsection{NEC violation}
Let us discuss the scalar field configuration and the violation of the NEC associated with the existence of the phantom scalar field. 
From Eq.~\eqref{eq_scal}, we see that,
at infinity, $\phi$ asymptotically approaches a constant value, 
and the energy-momentum tensor asymptotically vanishes.
From Eq.~\eqref{eq_stress_enegy}, the violation of NEC
can be explicitly checked as 
\begin{align}
\label{NEC_scal_abst}
      \Xi:=T_{\mu \nu}k^\mu k^\nu =-(k^\mu \partial_\mu\phi)^2 \le 0,
\end{align}
where $k^\mu$ is an arbitrary null vector field. 
In this paper, we investigate NEC violation with respect to the following specific null vector field:
\begin{align}
    k^\mu = \left(e^{-a},\frac{e^{q/2}}{\sqrt{p}},0,\omega e^{-a},\omega e^{-a}\right), 
\end{align}
indicating the null-zero-angular-momentum observer (ZAMO) frame, where $k^{\mu}$ satisfies $k^{\mu} k_{\mu} = 0, k_{\phi} = 0$, and $k_{\psi} = 0$.
$k^\mu$ is normalized by
\begin{align}
    -k^\mu u_\mu =1
\end{align}
with $u^\mu$ being the timelike ZAMO observer with the velocity vector $u^\mu=e^{-a}(1,0,0,\omega,\omega)$. 
For this null vector field, we find 
\begin{align}
\label{rela_Xi}
\Xi=-Q^2\frac{e^q}{p^3 h^3}.
\end{align}

\begin{figure}[htbp]
    \centering
        \includegraphics[width=0.4\textwidth]{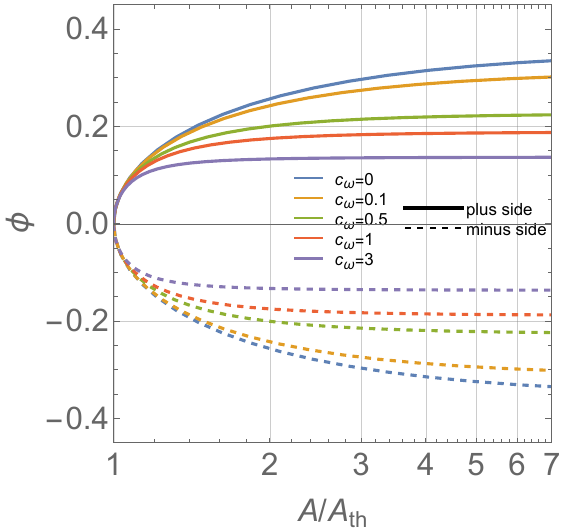} 
        \includegraphics[width=0.4\textwidth]{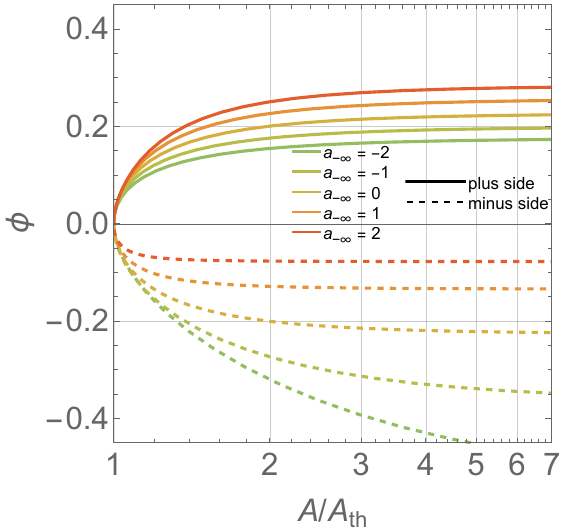} \\
        \includegraphics[width=0.4\textwidth]{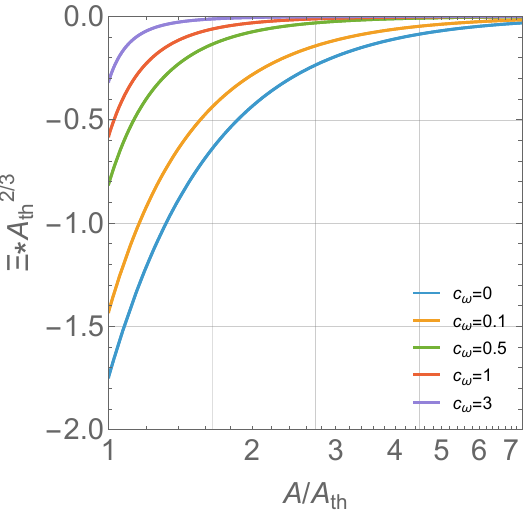} 
        \includegraphics[width=0.4\textwidth]{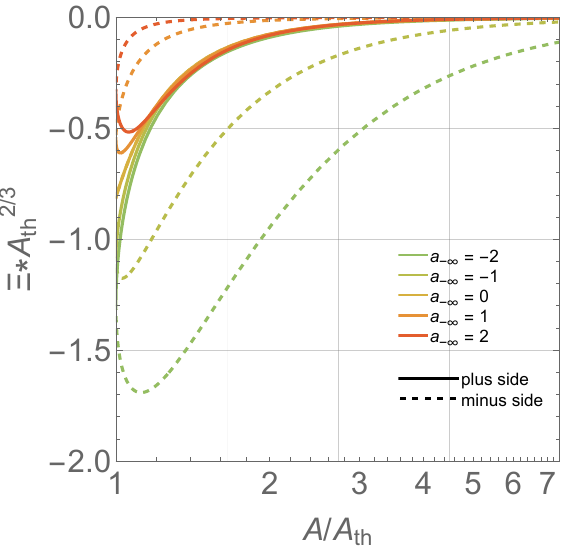} \\
    \caption{
               The solution of the scalar field $\phi$ (upper) and the null energy condition $\Xi$ (lower) as a function of area $A$ (with $G=1$). The left panels
               show the results for fixed $a_{-\infty}=0$ and $c_{\omega}=0, 0.1, 0.5, 1, 3$, while the right panels show those for fixed $c_{\omega}=0.5$ and $a_{-\infty}=-2, -1, 0, 1, 2$. For graphical convenience, the values of $\phi$ are shifted so that they become zero at the throat. 
               The solid and dashed curves describe the profiles in the region $l>l_{\rm th}$ and $l<l_{\rm th}$, respectively. }
    \label{fig_AvsphiNEC}
\end{figure}

We plot the values of $\phi$ and $\Xi$ in Fig.~\ref{fig_AvsphiNEC}.
The dependences on $c_{\omega}$ shown in the left columns reproduce the result reported in Ref.~\cite{Dzhunushaliev:2013jja}.
The negative values of $\Xi$ explicitly show the violation of the NEC.
However, increasing $c_{\omega}$ reduces the absolute value of $\Xi$, bringing it closer to $\Xi=0$, where the NEC is satisfied. Thus, increasing the angular momentum mitigates the violation of the null energy condition. 
We also find that, while $\Xi A_{\text{th}}^{2/3}$ reaches its minimum at the throat for the symmetric case ($a_{-\infty}=0$), it is not the case for the asymmetric wormhole ($a_{-\infty} \ne 0$). 

\begin{figure}[htbp]
  \centering
  \includegraphics[width=17cm]{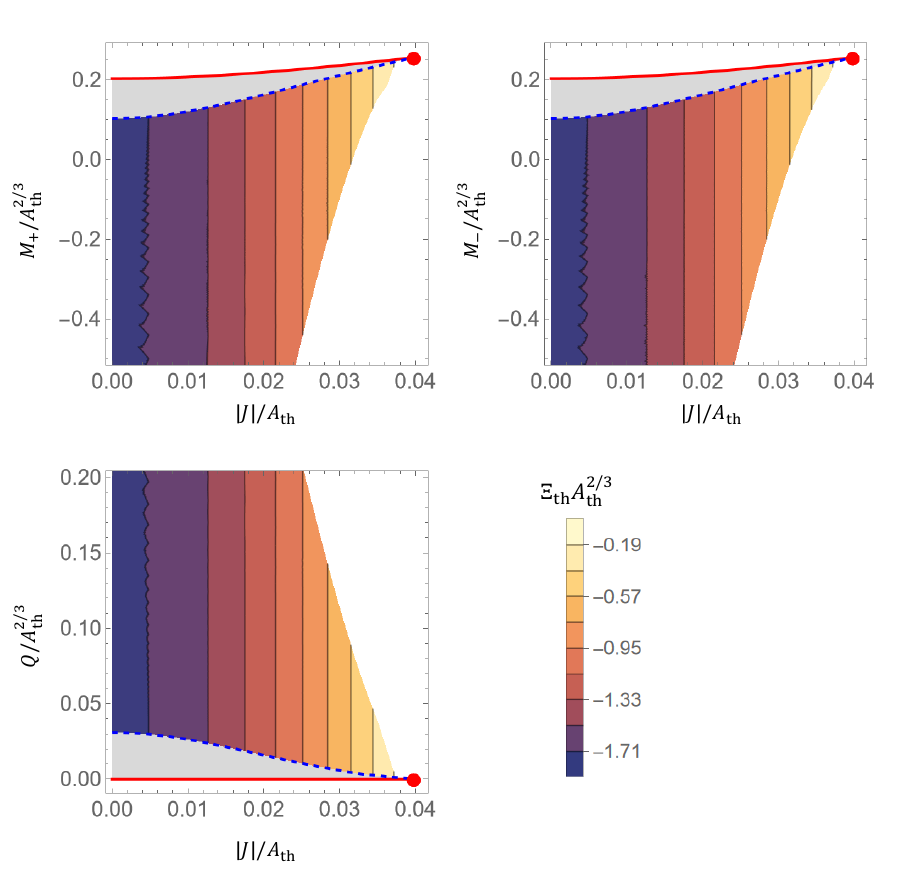}
  \caption{
  The contours of $\Xi_{\text{th}} A_{\text{th}}^{2/3}$, 
  the quantity evaluated at the throat,
  are overlaid on the phase diagrams shown in Fig.~\ref{fig_JvsQorM}, 
  plotted on the $|J|-M_+$ plane (upper left), the $|J|-M_-$ plane (upper right), and the $|J|-Q$ plane (lower left). 
  The red curve, blue-dashed curve, and gray-shaded region are identical to those in Fig.~\ref{fig_JvsQorM}, 
  representing the sequence of the Myers--Perry solutions, the limit curve of the wormhole solutions, and the gap region between the two curves, respectively.
  The color scale of the contours indicates the value of $\Xi_{\text{th}} A_{\text{th}}^{2/3}$: brighter colors correspond to weaker NEC violation. 
  }
  \label{fig_Plot_JvsMvsXi}
\end{figure}

We investigate the violation of the NEC 
at the throat 
given in Eq.~\eqref{rela_Xi} in the whole parameter region spanned by $c_\omega$ and $a_{-\infty}$ in the phase diagrams in Fig.~\ref{fig_Plot_JvsMvsXi}.  
It can be seen that the NEC contours are aligned vertically, indicating that they 
essentially depend only on the value of the angular momentum and are almost 
independent of $a_{-\infty}$.
This implies that, in rotating wormholes, the relaxation of the NEC violation is governed predominantly by the angular momentum parameter.

\section{Conclusions}
\label{sec:conclusion}

In this study, we analyzed rotating wormhole solutions with equal angular momenta in five dimensions, 
extending the previous work~\cite{Dzhunushaliev:2013jja}. In particular, we focused on the dependence on the asymmetry parameter $a_{-\infty}$, which originates from the mass difference between the two asymptotically flat regions connected by the wormhole. 
In Ref.~\cite{Dzhunushaliev:2013jja}, this parameter was restricted to the discrete values $-1, 0, 1$, but in the present work, we extended its range and systematically investigated its parameter dependence. 

We confirmed that the degree of violation of the null energy condition (NEC) decreases as the absolute value of the angular momentum increases.
We also found that, regardless of the value of the asymmetry parameter,
the violation of the NEC can be made arbitrarily small only in the rapid rotation limit, 
where the wormhole solution approaches the extremal Myers--Perry solution. 
Furthermore, by evaluating the violation of the NEC at the throat, we  
found that the degree of the violation depends 
predominantly on the angular momentum and is nearly independent of the asymmetry parameter. 
In other words, introducing and enlarging the asymmetry cannot mitigate the violation of the energy condition, and near-extremal rotation is essential for achieving a significant reduction of the NEC violation. 
We also found that the wormhole geometry approaches the extremal Myers--Perry black hole in the rapid rotation limit, while non-extremal Myers--Perry solutions cannot be obtained by any limit.

In this work, we provided new insights into the relationship between the asymmetry of rotating wormholes and the energy condition, and have comprehensively explored the parameter space that remained unexamined in previous studies~\cite{Dzhunushaliev:2013jja}.
The results obtained in this work may enable further investigations of rotating wormholes, in particular toward stability analysis. In fact, stability analyses have already been carried out for four-dimensional rotating wormholes \cite{Khoo:2024yeh, Azad:2024axu}.
However, in the rapid rotating limit, the problem becomes technically challenging due to the necessity of handling partial differential equations.
Therefore, it is essential to investigate the stability of five-dimensional rotating wormholes with equal angular momenta, such as those studied in the present work. 
Moreover, since we have also examined their relation to the extremal Myers--Perry black holes, our results might be applicable to the stability analyses of the extremal Myers--Perry black holes~\cite{Kodama:2009bf,Dias:2009iu,Dias:2010eu}.
Another direction would be to investigate particle orbits in these geometries, as discussed in~\cite{Kleihaus:2014dla, Chew:2016epf}.

\section*{Acknowledgment}
K.U. would like to take this opportunity to thank the “THERS Make New Standards Program for the Next
Generation Researchers” supported by JST SPRING, Grant Number JPMJSP2125. 
This work was partially supported by JSPS KAKENHI Grants Numbers JP23KK0048 (Y.K.), JP24KJ1223 (D.S.), JP25K07281 (C.Y.),  JP24K07027 (C.Y.), and JP21H05189 (D.Y.).

\appendix

\section{Metric components as functions of area}
\label{App:metric_function}

\begin{figure}[htbp]
    \centering
        \includegraphics[width=0.4\textwidth]{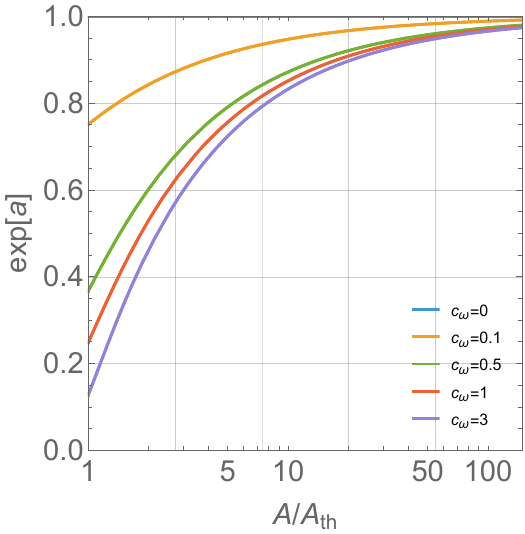} 
        \includegraphics[width=0.4\textwidth]{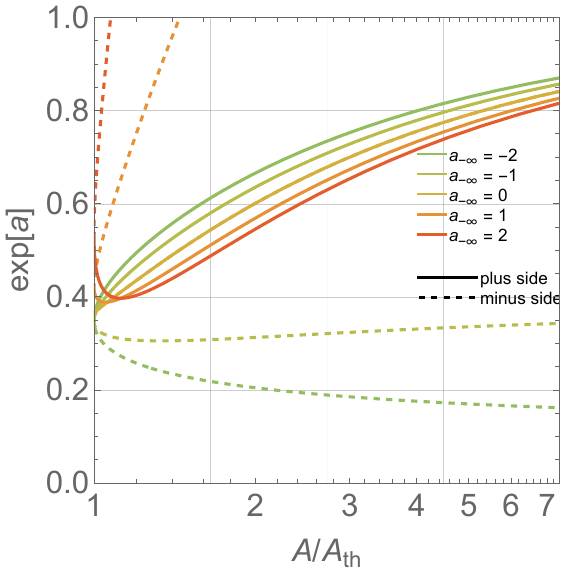} \\
        
        \includegraphics[width=0.4\textwidth]{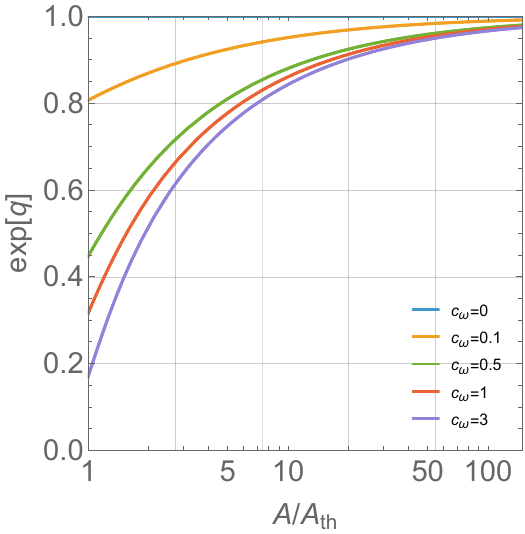} 
        \includegraphics[width=0.4\textwidth]{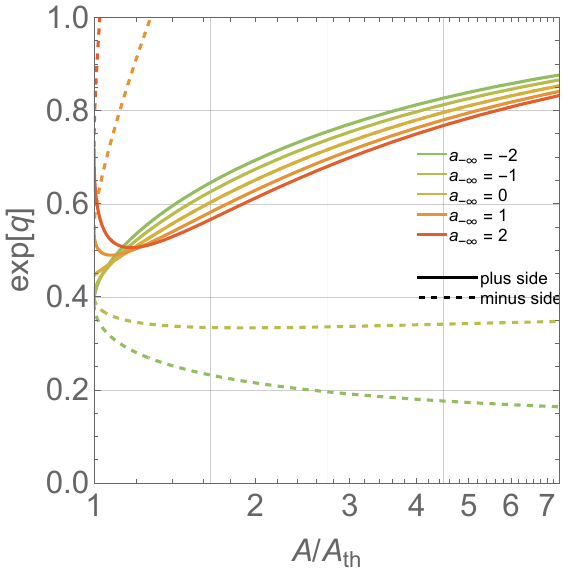} \\
        \caption{
               The solutions of the metric functions, $a$ and $q$ as functions of area $A$. The left 
               panels show the results for fixed $a_{-\infty}=0$ and $c_{\omega}=0, 0.1, 0.5, 1, 3$, while the right panels show those for fixed $c_{\omega}=0.5$ and $a_{-\infty}=-2, -1, 0, 1, 2$. The solid and dashed curves describe the profiles in the region $l>l_{\rm th}$ and $l<l_{\rm th}$, respectively. 
               }
    \label{fig_metfun0}
\end{figure}

\begin{figure}[htbp]
    \centering
        \includegraphics[width=0.4\textwidth]{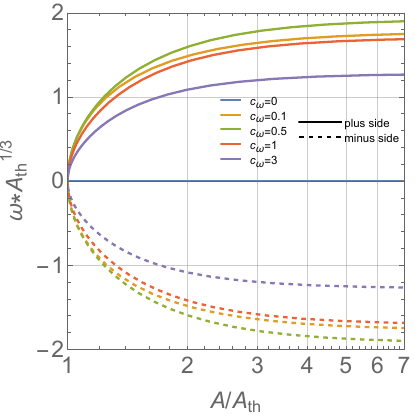} 
        \includegraphics[width=0.4\textwidth]{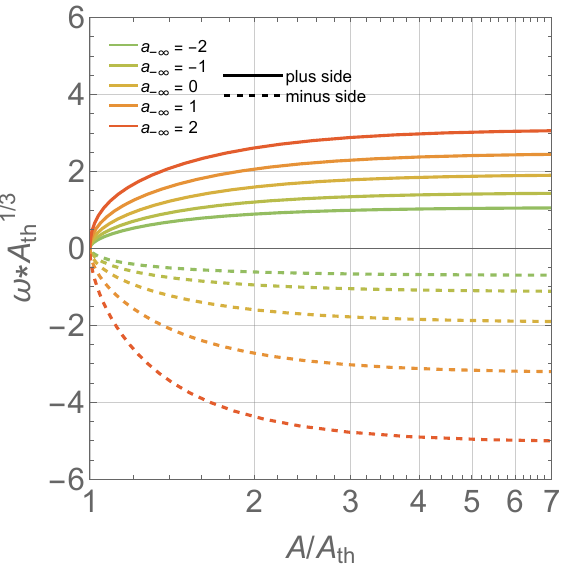} \\
    \caption{
               The solution of the metric function $\omega$ as a function of area $A$. The left panel
               shows the results for fixed $a_{-\infty}=0$ and $c_{\omega}=0, 0.1, 1, 3$, while the right 
               panel shows those for fixed $c_{\omega}=0.5$ and $a_{-\infty}=-2, -1, 0, 1, 2$. For graphical convenience, the values of $\omega$ are shifted so that they become zero at the throat. The solid and dashed curves describe the profiles in the region $l>l_{\rm th}$ and $l<l_{\rm th}$, respectively.
               }
    \label{fig_metfun}
\end{figure}

Here we show the explicit form of the metric functions $a$, $q$, and $\omega$ for $a_{-\infty}=0$ with various values of $c_\omega$, and for $c_{\omega}=0.5$ with various values of $a_{-\infty}$ 
as functions of the cross-sectional area $A$ in Fig.~\ref{fig_metfun}. 
The dependences on $c_{\omega}$ shown in the left columns are consistent with the results of Ref.~\cite{Dzhunushaliev:2013jja}.
It should be noted that, the radial coordinate $x$ is a two-valued quantity as a function of $A$ in general, because the decreasing cross-sectional area towards the throat starts to increase over the throat. 
Therefore, $a$, $q$, and $\omega$ are also considered as two-valued quantities as functions of $A$. 

For $a_{-\infty}=0$, which has the reflection symmetry with respect to the throat, $a$ and $q$ look single-valued functions since two curves overlap each other due to the reflection symmetry. 
On the other hand, $\omega$ becomes an odd function of $x$, and consequently, we find a symmetric figure with respect to the horizontal axis in the left lower panel in Fig.~\ref{fig_metfun}. 
As the angular momentum increases, $a$, $q$, and the absolute value of $\omega$ at a fixed position monotonically decrease.
For $a_{-\infty}\neq 0$, the functions $a$ and $q$ become two-valued (solid and dashed curves), and the figure for $\omega$ becomes asymmetric with respect to the horizontal line.

\section{The relation between wormhole solution and extremal Myers--Perry black hole solution}
\label{App:Rela_MPBH_WH_analy}
The Myers--Perry solution \cite{Myers:1986un, Myers:2011yc} represents the generalization of the Kerr solution in higher dimensions. This paper considers the only case of a five-dimensional spacetime with equal angular momenta. The metric of the Myers--Perry solution in this case is described as
\begin{align}
\label{met_MP}
    ds^2=&-dt^2+\frac{\Sigma}{\Delta}dr^2+\frac{\Sigma}{r^2}d\theta^2+\left( r^2+\alpha^2 \right)\sin^2\theta d\varphi^2+\left( r^2+\alpha^2 \right)\cos^2\theta d\psi^2\notag\\
    &+\frac{\mu r^2}{\Sigma} \left( dt-\alpha\sin^2\theta d\varphi - \alpha\cos^2\theta d\psi \right)^2 
\end{align}
with
\begin{align}
    \Sigma=r^2\left( r^2+\alpha^2 \right),\quad \Delta=\left( r^2+\alpha^2 \right)^2-\mu r^2.
\end{align}
The ranges of coordinates are defined as $0 < r < +\infty$, $0\le\theta<\pi/2$, $0\le\varphi<2\pi$, and $0\le\psi<2\pi$. 
$\mu$ and $\alpha$ are constant parameters that correspond to the mass and angular momentum, respectively.
In the case of the extremal Myers--Perry black hole, the parameter takes the value $\mu=4\alpha^2$.
Hereafter, we focus on the parameters with $\mu \geq 4 \alpha^2$, where the solution describes black hole spacetime.

We denote the horizon radius by $r_{\text{H}}$, which is determined as the largest real root of $\Delta=0$, namely
\begin{align}
(r_\text{H}^2+\alpha^2)^2-\mu r_{\text{H}}^2=0.
\end{align}
In the same manner as Eq.~\eqref{def_A}, the horizon area is then given by
\begin{align}
A_{\text{H}}=2\pi^2 \frac{(r_{\text{H}}^2+\alpha^2)^2}{r_{\text{H}}}.
\end{align}
The mass $M_{\text{MP}}$ and angular momentum $J_{\text{MP}}$ of the Myers–Perry solution are expressed as
\begin{align}
M_{\text{MP}}=\frac{3 \pi}{8G} \mu,
J_{\text{MP}}=\frac{\pi}{4G}\mu \alpha.
\end{align}
In the extremal case, where $\mu = 4\alpha^2$, the horizon radius reduces to $r_{\text{H}}=|\alpha|$. 
Substituting this relation into the above expressions, we obtain
\begin{align}
\label{exMP_mass}
\frac{GM_{\text{MP}}}{A_{\text{H}}^{2/3}} &= \frac{3}{8\pi^{1/3}} \simeq 0.256 , \\
\label{exMP_AM}
\frac{GJ_{\text{MP}}}{A_{\text{H}}} &= \frac{1}{8 \pi} \simeq 0.0398.
\end{align}

We examine whether the Myers--Perry solution can be accommodated within our framework through the coordinate transformation $r = r(l)$.
That is, we determine the functional forms of $a$, $q$, and $\omega$ that describe the Myers--Perry black hole and derive the possible constraints on the parameters,  
comparing the coefficients of the metrics~\eqref{met_MP} and \eqref{ansatz_rot_WH}.
Hereafter we express the specific functions for the Myers--Perry black hole as $a_{\text{MP}}, q_{\text{MP}}$, and $\omega_{\text{MP}}$. 
As the Myers--Perry black hole is a vacuum solution of the Einstein equations, the corresponding metric \eqref{ansatz_rot_WH} must also represent a vacuum solution. Therefore, it should correspond to the solution analyzed in the main section with $Q = 0$.

By comparing the $\theta\theta$, $\varphi\psi$, and $t\varphi$ components of each metric,  
we obtain
\begin{align}
\label{sol_omega_mp}
    \omega_\mathrm{MP}
    &=\frac{\mu \alpha}{(r^2+\alpha^2)^2+\mu \alpha^2},
\end{align}
as well as 
\begin{align}
    q_{\text{MP}} &= a_{\text{MP}} + \ln \left[
    \frac{\sqrt{(r^2 + \alpha^2)^2 + \mu \alpha^2}}{r^2 + \alpha^2}
    \right],
    \\
    (ph)_{\text{MP}} &= e^{a_{\text{MP}}} \sqrt{(r^2 + \alpha^2)^2 + \mu \alpha^2} .
\end{align}

Using these results together with the $tt$ component, we further find
\begin{align}
    a_{\text{MP}}
    &=\frac{1}{2}\ln{\left[\frac{\Delta}{(r^2+\alpha^2)^2+\mu \alpha^2}\right]}.
\end{align}
Substituting this expression back into the above relations, we finally obtain
\begin{align}
    \label{sol_q_mp}
    q_{\text{MP}}
    &=\frac{1}{2}\ln{\left[\frac{\Delta}{(r^2+\alpha^2)^2}\right]},\\
    \label{sol_ph_mp}
    (ph)_{\mathrm{MP}}
    &=\sqrt{\Delta}.
\end{align}
By comparing the $ll$ component of each metric, we obtain
\begin{align}
    \label{rela_rl}
    \frac{dr}{dl}&=\frac{p}{r}\sqrt{h},
\end{align}
where we have fixed the sign so that $r$ increases with $l$.
Integrating this equation, together with the concrete expressions for $p$ and $h$, Eqs.~\eqref{eq_sol_p} and \eqref{def h} with $(c_{1}, c_{2}) = (0, 1)$,
we can express the relation between two coordinates as
\begin{align}
    r^2&=l\sqrt{l^2+r_0^2}+c, \label{r as a function of l}
\end{align}
where $c$ is an integration constant. 
By substituting Eq.~\eqref{sol_omega_mp} and Eq.~\eqref{rela_rl} into 
Eq.~\eqref{eq_Einstein_omega}, we find
\begin{align}
    c_{\omega}=-4\mu \alpha.
\end{align}

Let us check the consistency 
of the formulas obtained here.
By substituting the expression \eqref{r as a function of l} into \eqref{sol_ph_mp}, we obtain
\begin{align}
 - (c+\alpha^2)^2 + c \mu + \frac{r_{0}^2}{4}  - 2 \left( c +\alpha^2 - \frac{1}{2} \mu \right) l \sqrt{l^2 + r_{0}^2} = 0,
\end{align}
for any $l$. Therefore, the constants must satisfy
\begin{align}
c = \frac{1}{2} (\mu - 2 \alpha^2), 
\end{align}
and 
\begin{align}
r_{0}^4 = \mu (4 \alpha^2 - \mu). \label{r4 =}
\end{align}
While the left hand side is non-negative, the right hand side is non-positive because of $\mu \geq 4 \alpha^2$. Therefore, Eq.~\eqref{r4 =} can be satisfied when
\begin{align}
r_{0} = 0,\qquad \mu = 4 \alpha^2. \label{con_extremal}
\end{align}

Since Eq.~\eqref{con_extremal} is the condition of the extremal Myers--Perry solution, the metric ansatz \eqref{ansatz_rot_WH} includes the Myers--Perry solution only in the extremal case. 

\section{Static solution}
\label{app:static_solution}
Here, we describe static and spherically symmetric wormhole solutions in five dimensions including the asymmetric configurations. 
These symmetries require the conditions $c_\omega = 0$ and $a = q$.
Consequently, the differential equations \eqref{eq_Einstein_a} and \eqref{eq_Einstein_q}, which determine $a(l)$ and $q(l)$, reduce to
\begin{align}
\label{solnonrot}
    (ph^{3/2}a')' = 0.
\end{align}
This equation can be solved analytically as follows:
\begin{align}
\label{solanonrot}
    a(l) = -\frac{2\rho}{r_0^2} \arctan{\left[\frac{1 - l/\sqrt{r_0^2 + l^2}}{1 + l/\sqrt{r_0^2 + l^2}}\right]} + c_0,
\end{align}
where $ \rho $ and $ c_0 $ are integration constants. Imposing the asymptotic condition $ a_{+\infty} = 0 $, we obtain
\begin{align}
    c_0 = 0.
\end{align}

Expanding Eq.~\eqref{solanonrot} in terms of $ 1/l $, we obtain:
\begin{align}
a(l) = 
\begin{cases}
    -\frac{\rho}{2}\frac{1}{l^2} + O(l^{-4}) & (l \to +\infty),  \\
    -\frac{\rho \pi}{r_0^2} + \frac{\rho}{2}\frac{1}{l^2} + O(l^{-4}) & (l \to -\infty).
\end{cases}
\end{align}
Thus, the constant $ a_{-\infty} $ can be expressed in terms of $ \rho $ as:
\begin{align}
\label{muainf}
    a_{-\infty} = -\frac{\pi\rho }{r_0^2}.  
\end{align}
The area of the static case is
\begin{align}
    \label{rela_A}
    A&=2\pi^2(ph)^{3/2} \exp\left[-\frac{3a_{-\infty}}{\pi} \arctan{\left(\frac{1 - l/\sqrt{r_0^2 + l^2}}{1 + l/\sqrt{r_0^2 + l^2}}\right)} \right]. 
\end{align}
Then the value of the coordinate $l$ at the throat is given by
\begin{align}
    l_{\text{th}}=\frac{-a_{-\infty}r_0}{\abs{a_{-\infty}}}\sqrt{\frac{-
    1+\sqrt{1+(a_{-\infty}/\pi)^2}}{2}}.
\end{align}
The area of the throat is
\begin{align}
    A_{\text{th}}=r_0^3\sqrt{\frac{\pi}{2}}(a_{-\infty}^2+\pi^2)^{3/4}\exp \left[-\frac{3a_{-\infty}}{\pi} \arctan{\left(\sqrt{1+\left(\frac{a_{-\infty}}{\pi}\right)^2}+\frac{a_{-\infty}}{\pi}\right)} \right].
\end{align}

By using Eqs.~\eqref{solanonrot} and  \eqref{rela_M_a2}, the mass can be related to $a_{-\infty}$ as follows:
\begin{align}
    M_\pm 
    = \mp\frac{3r_0^2 e^{-a_{\pm \infty}}}{8 G} a_{-\infty}.
\end{align}
From Eq.~\eqref{rela_Q_q4}, the scalar charge $Q$ with $a_{-\infty}$ is given by 
\begin{align}
Q^2 =  \frac{3}{16 \pi G}r_0^4\left(1 + \left(\frac{a_{-\infty}}{\pi}\right)^2\right).
\end{align}
Then, the asymptotic expansion of the normalized mass $M_+$ and scalar charge for $a_{-\infty} \to -\infty$ is given by
\begin{align}
   \frac{GM_+}{A_{\rm th}^{2/3}} &= \frac{3}{4\cdot2^{2/3}e\pi^{1/3}}+O\left(\left(\frac{1}{a_{-\infty}}\right)^2\right)\approx 0.119+O\left(\left(\frac{1}{a_{-\infty}}\right)^2\right), \\
    \frac{\sqrt{G}Q}{A_{\rm th}^{2/3}} &= \frac{\sqrt{3}}{2\cdot2^{2/3}e\pi^{11/6}}+O\left(\left(\frac{1}{a_{-\infty}}\right)^2\right)\approx 0.0246+O\left(\left(\frac{1}{a_{-\infty}}\right)^2\right).
\end{align}
Similarly, we obtain
\begin{align}
   \frac{GM_-}{A_{\rm th}^{2/3}} &= \frac{3}{4\cdot2^{2/3}e\pi^{1/3}}+O\left(\left(\frac{1}{a_{-\infty}}\right)^2\right)\approx 0.119+O\left(\left(\frac{1}{a_{-\infty}}\right)^2\right)
\end{align}
for $a_{-\infty}\rightarrow\infty$.
These values are reflected in Fig.~\ref{fig_JvsQorM}.


\providecommand{\href}[2]{#2}\begingroup\raggedright\endgroup

\bibliographystyle{jhep}

\end{document}